\documentclass[journal=jpclcd,manuscript=letter]{achemso}
\usepackage{chemformula}
\usepackage[T1]{fontenc} 
\usepackage{xcolor,color}
\usepackage{verbatim}
\usepackage{float}
\usepackage{mathrsfs}
\usepackage{amsmath}
\usepackage{amssymb}
\usepackage{graphics}
\usepackage{graphicx}
\usepackage{physics}
\usepackage{hyperref}
\hypersetup{
    colorlinks=true,
    citecolor=red,
    linkcolor=blue,
    filecolor=magenta,      
    urlcolor=blue,
    pdfpagemode=FullScreen,
    }

\DeclareMathAlphabet{\mathpzc}{OT1}{pzc}{m}{it}

\newcommand{\mA}{\mathrm{\scriptscriptstyle A}}
\newcommand{\mD}{\mathrm{\scriptscriptstyle D}}
\newcommand{\nequil}{\mathrm{\scriptscriptstyle (neq)}}

\newcommand{\cminv}{\mathrm{cm}^{-1}}
\newcommand{\mRC}{\mathrm{RC}}

\usepackage[symbol]{footmisc}

\usepackage[normalem]{ulem}

\newcommand{\CPTG}{
Chemical Physics Theory Group, Department of Chemistry,
and Center for Quantum Information and Quantum Control,
University of Toronto, Toronto, Ontario M5S 3H6, Canada}

\author{Leonardo F. Calder\'on}
\email{leonardo.calderon@utoronto.ca}
\affiliation{\CPTG}
\author{Paul Brumer}
\email{paul.brumer@utoronto.ca}
\affiliation{\CPTG}
\title[]
{Frequency-Dependent Vibronic Effects in Steady State Energy Transport}
\date{\today}

\begin{document}

\begin{tocentry}
\includegraphics[scale=0.47]{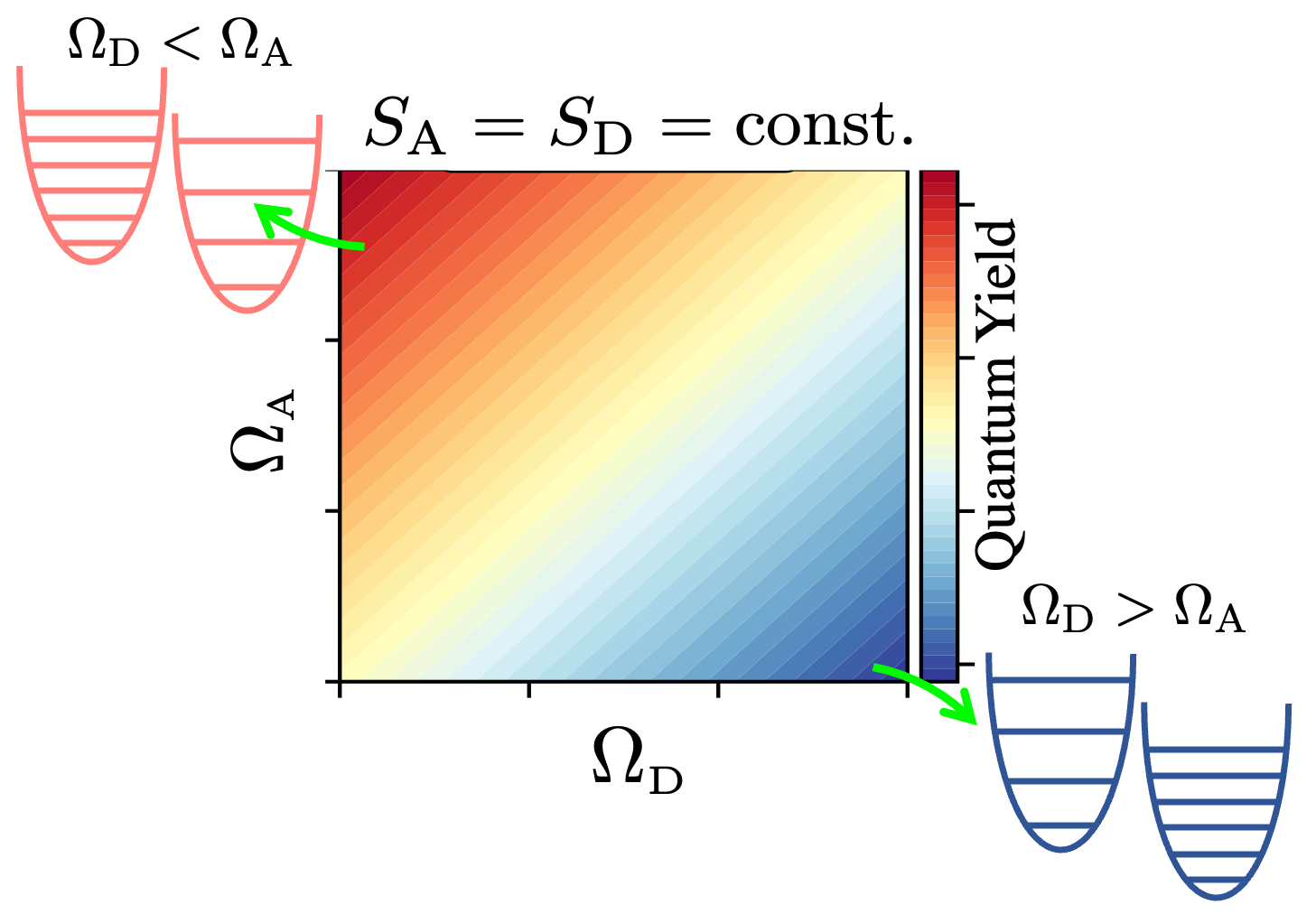}
\end{tocentry}

\begin{abstract}
The interplay between electronic and intramolecular high-frequency 
vibrational degrees of freedom is ubiquitous in natural light-harvesting 
systems. 
Recent studies have indicated that an intramolecular vibrational 
donor-acceptor frequency difference can enhance energy transport. 
Here, we analyze the extent to which different intramolecular donor-acceptor 
vibrational frequencies affect excitation energy transport in equilibrium 
(coherent light excitation) and the more natural nonequilibrium steady state 
(incoherent light excitation) configurations. 
It is found that if the Huang-Rhys factors remain 
constant, the acceptor population increases when the intramolecular 
vibrational frequency of the acceptor exceeds that of the donor.
The increase in the acceptor population due to the vibrational frequency 
difference is higher for higher values of the Huang-Rhys factors or the 
vibronic coupling strengths.
However, the nonequilibrium steady state results show that the vibrational 
donor-acceptor frequency difference does not significantly enhance energy 
transport in the natural scenario of incoherent light excitation and under 
biologically relevant parameters.
Insight about a potential mechanism to optimize energy transfer in the NESS 
based on increasing the harvesting time at the reaction center is analyzed.
\end{abstract}

\textbf{\textit{Introduction}}.\textemdash
Excitation energy transport in natural light-harvesting systems (LHS), e.g., 
photosynthetic pigment-protein complexes, has been intensively studied  
with an aim of extracting design principles applicable to artificial 
(human-made) light-harvesting systems \cite{SF&11,RNG17,CC&20,Man20}. 
It is worth noting that the interaction between molecular electronic and 
vibrational degrees of freedom influences the energy transfer in LHS 
\cite{PC&10,WM11,BVA12,CK&12,KO&12,CC&13,TPJ13,CP&13,FP&14,BVA14,HJ&14,OO14,
RA&14,DW&15,NN&15,SI&15,MS&16,DM&16,YH&19,CC&20,AY&20,HL&21,PN&22}. 
The vibrational degrees of freedom can be classified as inter- and 
intramolecular vibrational modes.
The intermolecular vibrational modes are usually modeled as a low-frequency 
phonon bath characterized by a spectral density.
The intramolecular vibrational modes are considered as discrete underdamped 
vibrations of high-frequency.
Experimental spectra results of photosynthetic pigment-protein complexes 
indicate the presence of several intramolecular vibrational modes with 
different frequencies and strengths coupling to the electronic degrees of 
freedom \cite{WP&00,RF07,CK&12,HJ&14,KL&20}.
Despite different high-energy vibrational frequency modes in the spectra, 
most theoretical models of light harvesting systems accounting for vibronic 
effects consider couplings to vibrational modes of the same frequency.

Recently, it has been reported that the energy 
transfer quantum yield in a prototype photosynthetic dimer is enhanced 
when the donor is coupled to a vibrational mode whose frequency is larger 
than that on the vibrational mode coupled to the acceptor \cite{DN&20}. 
In this donor-acceptor energy transfer \cite{DN&20}, it was assumed that the 
entire population is initially in the donor chromophore, representing a 
scenario commonly considered under coherent light excitation conditions. 
These laboratory-designed/controlled conditions contrast with those in  
nature, where the natural light-harvesting systems are continuously 
illuminated by incoherent radiation (sunlight) \cite{JB91,MV10,BS12,PB12,KYR13,PBB17,Bru18,TK20,CP20,DB21,CCB23}, leading to a nonequilibrium
steady state (NESS) \cite{TB18,Bru18,CCB23,DB21,CB20,JB20}.
In this Letter, we examine the conditions that lead to an increase in the 
quantum yield for both equilibrium and nonequilibrium steady states for 
coherent (laboratory) and incoherent (natural) light excitation, considering 
different intramolecular vibrational donor-acceptor frequencies and different 
Huang-Rhys factors for a prototypical photosynthetic dimer.

It is important to note that for energy transfer technologies even small
improvements in energy transfer, e.g. $5\%$, can be significant.
Here we show that if such improvements are seen in equilibrium they are
washed out in the NESS regime. 
Furthermore,  in the particular case of biophysical systems we show that, 
contrary to some suggestions in the literature, donor-acceptor frequency 
differences do not significantly affect energy transfer.

We consider a prototype photosynthetic electronic dimer, adopting 
parameters typical of the Fenna-Matthews-Olson (FMO) complex 
\cite{WP&00,RF07,CK&12,HJ&14,KL&20}.
The dimer is immersed in a protein-solvent environment modeled 
as a vibrational thermal bath at $300\,\mathrm{K}$, composed of intra- and 
intermolecular vibrational modes and describe the incoherent light as a 
blackbody radiation thermal bath at $5600\,\mathrm{K}$ \cite{PB13,PBB17}.
Additionally, exciton recombination and harvesting are considered.  
We solve for the equilibrium and nonequilibrium steady state using the 
numerically exact hierarchical equations of motion (HEOM) method 
\cite{TK89,Tan90,IT05,IF09,Tan20,LR&20}.
The steady state acceptor population is regarded as the criterion to 
evaluate the energy transfer efficiency from donor to acceptor.

In this letter, we analyze the steady state acceptor population
for different intramolecular vibrational donor-acceptor 
frequencies, Huang-Rhys factors, and vibronic coupling strengths in an 
equilibrium configuration (dimer + vibrational bath only), and in the 
nonequilibrium configuration (dimer + vibrational bath + radiation bath +
exciton recombination and harvesting).
In doing so, we demonstrate, for example, and as noted above, that 
in the natural scenario of incoherent light excitation and under 
biologically relevant parameters, the vibrational donor-acceptor 
frequency difference does not significantly enhance energy transport.
In addition, we correct recently published erroneous results\cite{DN&20}, 
for the equilibrium steady state case, which states that the acceptor 
population increases when the intramolecular vibrational frequency of the donor 
is larger than at the acceptor for constant Huang-Rhys factors.
Furthermore, we find that if the vibronic coupling strengths are fixed,  
the acceptor population will increase when the vibrational frequency of the 
donor is larger than at the acceptor.

\textbf{\textit{Model}}.\textemdash
Consider a prototype photosynthetic complex modeled as a dimer 
(donor-acceptor (\textrm{D}-\textrm{A}) configuration) immersed within a 
protein-solvent environment.
We assume an electronic dimer model representing the system of interest 
using open quantum system methodology and described through the Frenkel 
Hamiltonian  
\begin{equation}
\label{eq:Hs_ele}
\hat{H}_{\mathrm{Sys}} =\,
\varepsilon_{\mD}\hat{\varphi}_{\mD}^{+}\hat{\varphi}_{\mD}^{-}
+ \varepsilon_{\mA}\hat{\varphi}_{\mA}^{+}\hat{\varphi}_{\mA}^{-}
+V_{\mD\mA}\left(\hat{\varphi}_{\mD}^{+}\hat{\varphi}_{\mA}^{-} 
                +\hat{\varphi}_{\mA}^{+}\hat{\varphi}_{\mD}^{-}\right).
\end{equation}
The two first terms of the right-hand side of Eq.~\ref{eq:Hs_ele} account
for the donor (acceptor) excited state electronic energies 
$\varepsilon_{\mD}$ ($\varepsilon_{\mA}$), and the last term for the 
excitonic coupling $V_{\mD\mA}$.
The electronic excitation creation (annihilation) operators
$\hat{\varphi}_{\mD,\mA}^{+}\left(\hat{\varphi}_{\mD,\mA}^{-}\right)$  
are defined by
$|\varepsilon_{\mD,\mA}\rangle = 
\hat{\varphi}_{\mD,\mA}^{+}|g_{\mD,\mA}\rangle,\,|g_{\mD,\mA}\rangle =
\hat{\varphi}_{\mD,\mA}^{-}|\varepsilon_{\mD,\mA}\rangle$.
We consider the two-level approximation on each site, where only the electronic 
ground state $|g_{\mD,\mA}\rangle$ and the electronic first excited state 
$|\varepsilon_{\mD,\mA}\rangle$ are taken into account.
The eigenstates of system 
$\hat{H}_{\mathrm{Sys}}|e_{i}\rangle=E_{e_{i}}|e_{i}\rangle$ 
denote the exciton basis
$\{|e_{+}\rangle,|e_{-}\rangle\}$, where
the exciton states can be defined as linear superpositions of the
electronic site states 
$|e_{i}\rangle=\sum_{j}^{\mD,\mA}c_{ij}|\varepsilon_{j}\rangle$, 
with energies 
$E_{e_\pm} = \frac{\varepsilon_{\mA}+\varepsilon_{\mD}}{2}
\pm\frac{1}{2}\sqrt{\Delta\varepsilon^{2}+4V_{\mD\mA}^{2}}$, and 
where $\Delta\varepsilon=\varepsilon_{\mD}-\varepsilon_{\mA}$ is
the site energy difference.

The protein-solvent environment is considered as a collection of harmonic 
vibrational modes.
This vibrational bath is composed of both intra- and intermolecular vibrational 
modes. 
The Hamiltonian for the vibrational bath reads
\begin{equation}
\label{eq:HVB}
\hat{H}_{\mathrm{VB}}=  
\hbar\Omega_{\mD}\hat{\mathfrak{b}}_{\mD}^{\dagger}\hat{\mathfrak{b}}_{\mD}
+ \hbar\Omega_{\mA}\hat{\mathfrak{b}}_{\mA}^{\dagger}\hat{\mathfrak{b}}_{\mA}
+ \sum_{i}^{\mD,\mA}\sum_{j}\hbar\omega_{j}^{(i)}
   \hat{b}_{j}^{(i)\dag}\hat{b}_{j}^{(i)}.
\end{equation}
We consider only one intramolecular vibrational on each site.
The creation (annihilation) operators for the intramolecular vibrational 
modes of frequencies $\Omega_{\mD}$ and $\Omega_{\mA}$ are denoted by 
the calligraphic letters 
$\hat{\mathfrak{b}}_{\mD,\mA}^{\dagger},\,(\hat{\mathfrak{b}}_{\mD,\mA})$. 
The creation (annihilation) operators for the intermolecular vibrational 
modes of frequencies $\omega_{l}^{(i)}$  are denoted by 
$\hat{b}_{l}^{(i)\dag},\,(\hat{b}_{l}^{(i)})$.

The interaction between the system and the vibrational bath is given 
by the Hamiltonian
\begin{equation}
\label{eq:Hs-VB}
\begin{split}
\hat{H}_{\mathrm{Sys-VB}} = & \,
\hbar\mathcal{G}_{\mD}\hat{\varphi}_{\mD}^{+}
\hat{\varphi}_{\mD}^{-}\left(\hat{\mathfrak{b}}_{\mD}^{\dagger}
+\hat{\mathfrak{b}}_{\mD}\right)
+\hbar\mathcal{G}_{\mA}\hat{\varphi}_{\mA}^{+}
 \hat{\varphi}_{\mA}^{-}\left(\hat{\mathfrak{b}}_{\mA}^{\dagger}
 +\hat{\mathfrak{b}}_{\mA}\right)
\\ &
+\sum_{i}^{\mD,\mA}\sum_{j}\hbar g_{j}^{(i)}
\hat{\varphi}_{i}^{+}\hat{\varphi}_{i}^{-}
\left(\hat{b}_{j}^{(i)}+{\hat{b}_{j}}^{(i)\dag}\right),
\end{split}
\end{equation}
where $\mathcal{G}_{\mD}=\sqrt{S_{\mD}}\Omega_{\mD}$ 
($\mathcal{G}_{\mA}=\sqrt{S_{\mA}}\Omega_{\mA}$) denotes the 
coupling between the electronic donor (acceptor) state and the intramolecular 
vibrational mode of frequency $\Omega_{\mD}$ ($\Omega_{\mA}$) (referred to 
below as vibronic coupling), where $S_{\mD}$ ($S_{\mA}$) is the 
dimensionless Huang-Rhys factor at the donor (acceptor). 
In Eq.~\ref{eq:Hs-VB}, $g_{j}^{(\mD)}$ ($g_{j}^{(\mA)}$) represents the 
coupling between the electronic donor (acceptor) state and the 
$j^{\mathrm{th}}$ intermolecular vibrational mode.
Therefore, the global Hamiltonian for the prototype photosynthetic complex 
is given by 
$\hat{H}=
\hat{H}_{\mathrm{Sys}}+\hat{H}_{\mathrm{Sys-VB}}+\hat{H}_{\mathrm{VB}}$.

All the information about the vibrational bath is encoded in the spectral 
density of the donor and acceptor
\begin{equation}
\label{eq:SD_DA}
\begin{split}
J_{\mD,\mA}(\omega) & = \,
J_{\mathrm{intra}}(\omega)+ J_{\mathrm{inter}}(\omega) 
\\ &
=\frac{4\Lambda_{\mD,\mA}\Gamma\Omega_{\mD,\mA}^2\omega}
{(\Omega_{\mD,\mA}^2-\omega^2)^2 + 4\Gamma^2\omega^2} 
+
\frac{2\lambda\gamma\,\omega}{\hbar(\omega^2+\gamma^2)}.
\end{split}
\end{equation}
The first term in Eq.~\ref{eq:SD_DA} corresponds to the contribution of the 
intramolecular vibrational mode through an underdamped Brownian oscillator 
spectral density. 
The second term accounts for the intermolecular 
vibrational modes, corresponding to a low-frequency vibrational bath 
described by a Drude-Lorentz spectral density. 
Here, $\Lambda_{\mD,\mA}= S_{\mD,\mA}\Omega_{\mD,\mA} = 
\sqrt{S_{\mD,\mA}} \mathcal{G}_{\mD,\mA}$ is the reorganization 
energy and $\Gamma$ is the peak width (cut-off frequency).
We assume independent baths on each site at temperature 
$T^{\mathrm{VB}}=300\,\mathrm{K}$.

\textbf{\textit{Equilibrium steady state configuration}}.\textemdash
We analyze the effect of the vibrational bath described by 
Eqs.~\ref{eq:HVB}--\ref{eq:SD_DA} on the energy transfer from the donor to 
the acceptor by examining acceptor population changes due to vibronic 
effects in the steady state. 
The entire population is initially assumed to be in the donor 
chromophore $\rho_{\mD\mD}(t=0)=1$. 
This condition is commonly considered under coherent light excitation 
conditions.
Consistent with the literature in this scenario, the steady state corresponds 
to a thermal equilibrium state, where the acceptor population $\rho_{\mA\mA}$ 
quantifies the quantum yield.
Specifically, we consider different intramolecular vibrational donor 
and acceptor frequencies for several fixed Huang-Rhys factors and 
vibronic coupling strengths.
Other physical effects, such as incoherent light excitation, exciton 
recombination and harvesting, are analyzed below.
We assume parameters characteristic of prototypical photosynthetic 
complexes, such as the Fenna-Matthews-Olson (FMO) complex, i.e., 
$\Delta\varepsilon = 100\,\cminv$, with $\varepsilon_{\mD} > \varepsilon_{\mA}$, 
$V_{\mD\mA}=50\,\cminv$, the excitonic energy splitting is given by 
$\Delta E_{e} = 141.4\,\cminv$, and for the vibrational bath 
$\lambda=50\,\cminv$, $\gamma=200\,\cminv$, $\Gamma=10\,\cminv$, and
$T^{\mathrm{VB}}=300\,\mathrm{K}$.
%

\begin{figure*}[t]
\includegraphics[width=0.32\linewidth]{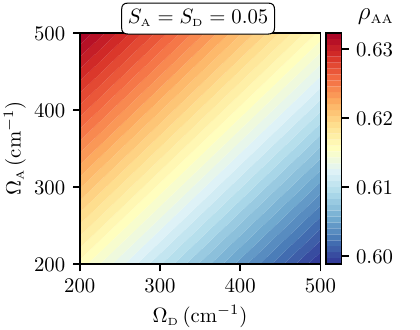} 
\includegraphics[width=0.32\linewidth]{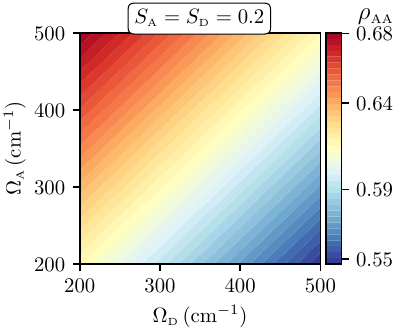} 
\includegraphics[width=0.32\linewidth]{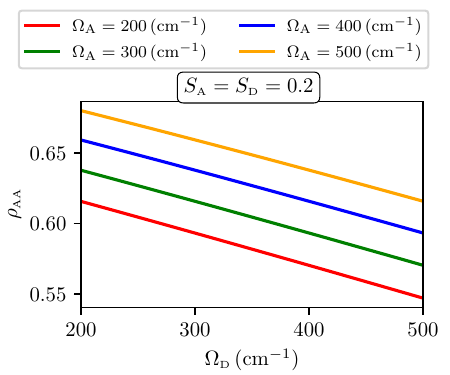}
\caption{
Equilibrium steady state acceptor population $\rho_{\mA\mA}$ as a 
function of the intramolecular vibrational frequencies at the donor 
$\Omega_{\mD}$ and the acceptor $\Omega_{\mA}$ considering 
different fixed Huang-Rhys factors $S_{\mA}=S_{\mD}=\{0.05,0.2\}$. 
The vibrational bath parameters are $\lambda=50\,\cminv$, 
$\gamma=200\,\cminv$, $\Gamma=10\,\cminv$ , and 
$T^{\mathrm{VB}}=300\,\mathrm{K}$.
}
\label{fig:popA_SS}
\end{figure*}
%
Figure~\ref{fig:popA_SS} depicts the thermal equilibrium 
acceptor population $\rho_{\mA\mA}$ (quantum yield) as a 
function of the intramolecular vibrational donor $\Omega_{\mD}$ and 
acceptor $\Omega_{\mA}$ frequencies for two constant 
Huang-Rhys factors $S_{\mA}=S_{\mD}=\{0.05,0.2\}$.
The intramolecular vibrational frequencies satisfy 
$\Omega_{\mD},\Omega_{\mA}>\Delta\varepsilon,\Delta E_e,V_{\mD\mA}$.
Figure~\ref{fig:popA_SS} shows that  
$\rho_{\mA\mA}$ increases when the intramolecular vibrational frequency at 
the acceptor is larger than at the donor $\Omega_{\mA}>\Omega_{\mD}$ and 
decreases when $\Omega_{\mA}<\Omega_{\mD}$.
For constant Huang-Rhys factors 
$S_{\mA}=S_{\mD}$, the thermal 
equilibrium acceptor population decreases linearly\footnote{The decrease in 
$\rho_{\mA\mA}$ is no longer linear for large values of the Huang-Rhys 
factors ($S_{\mA}=S_{\mD}\geq1$).} with $\Omega_{\mD}$ 
when $\Omega_{\mA}$ is constant, i.e., 
$\rho_{\mA\mA}(\Omega_{\mD};\Omega_{\mA},S_{\mA}=S_{\mD})\propto-\Omega_{\mD}$
(see the right panel in Figure~\ref{fig:popA_SS}). 
In this case, $\rho_{\mA\mA}$ becomes larger when 
$\Omega_{\mA} > \Omega_{\mD}$ because the vibronic 
coupling in the acceptor is larger than in the donor, i.e., 
$G_{\mA}>G_{\mD}$.
That means, for instance, when $\Omega_{\mA}$ is constant ($G_{\mA}$ is also 
constant), if $\Omega_{\mD}$ 
increases, then $G_{\mD}$ increases, since $S_{\mD}$ is constant, which leads 
to lower eigenenergies at the 
donor and, therefore larger donor populations and lower acceptor populations 
(thermal equilibrium distribution).

Figure~\ref{fig:popA_SS} shows that when $S_{\mA}$ and $S_{\mD}$ increase, 
the maximum and minimum of $\rho_{\mA\mA}$ increases and decreases, 
respectively.
The latter is a consequence of the increase in the 
vibronic coupling, which leads to lower and higher eigenenergies. 
For relevant biological values of the Huang-Rhys factors:   
$S_{\mA}=S_{\mD}=0.05$, $\rho_{\mA\mA}$ changes by up to $\sim 5\%$.
Hence, devices designed for enhanced energy transfer would benefit from 
larger Huang-Rhys factors.
The changes in $\rho_{\mA\mA}$ 
are marginal when $\Omega_{\mA}=\Omega_{\mD}$ ($\lesssim 0.1\%$), 
i.e., values along the diagonal. 
Note that the results for $S_{\mA}=S_{\mD}=0.2$ (middle panel in 
Figure~\ref{fig:popA_SS}) were recently obtained incorrectly \cite{DN&20}, 
being opposite to the results presented in this Letter. 
%

\begin{figure*}[t]
\includegraphics[width=0.32\linewidth]{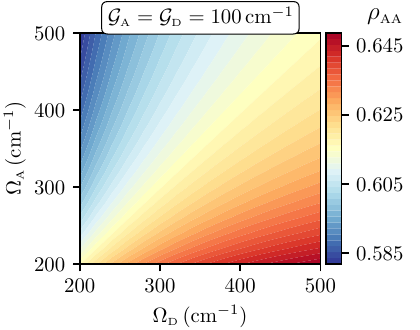}
\includegraphics[width=0.32\linewidth]{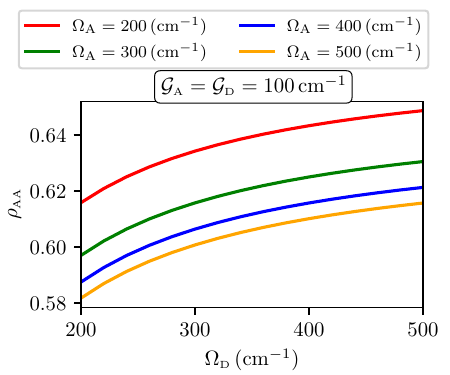} 
\includegraphics[width=0.32\linewidth]{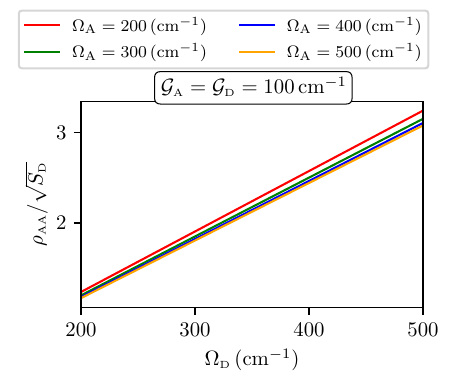}
\caption{
Equilibrium steady state acceptor population $\rho_{\mA\mA}$ as a function of 
the intramolecular vibrational frequencies at the donor 
$\Omega_{\mD}$ and the acceptor $\Omega_{\mA}$ considering 
fixed vibronic coupling strengths  
$\mathcal{G}_{\mA}=\mathcal{G}_{\mD}=100\,\cminv$. 
The vibrational bath parameters are $\lambda=50\,\cminv$, 
$\gamma=200\,\cminv$, $\Gamma=10\,\cminv$ , and 
$T^{\mathrm{VB}}=300\,\mathrm{K}$.
}
\label{fig:popA_GG}
\end{figure*}
%
Figure~\ref{fig:popA_GG} shows the thermal equilibrium 
acceptor population $\rho_{\mA\mA}$ (quantum yield) as a 
function of $\Omega_{\mD}$ and $\Omega_{\mA}$ for a constant 
vibronic coupling strength $G_{\mA}=G_{\mD}=100\,\cminv$. 
Figure~\ref{fig:popA_GG} shows that $\rho_{\mA\mA}$ 
increases when the intramolecular vibrational frequency at 
the donor is larger than at the acceptor $\Omega_{\mA}<\Omega_{\mD}$ 
and decreases when $\Omega_{\mA}>\Omega_{\mD}$ 
($\rho_{\mA\mA}$ changes by up to $\sim 11\%$).
This contrasts with the case of constant Huang-Rhys factors discussed 
above.
For constant vibronic coupling strengths 
$\mathcal{G}_{\mA} = \mathcal{G}_{\mD}$, the thermal equilibrium
acceptor population increases proportionally to 
$\sqrt{S_{\mD}}\,\Omega_{\mD}$ when $\Omega_{\mA}$ is constant, i.e., 
$\rho_{\mA\mA}(\Omega_{\mD};\Omega_{\mA},\mathcal{G}_{\mA} = 
\mathcal{G}_{\mD})\propto\sqrt{S_{\mD}}\,\Omega_{\mD}$  
(see the middle panel in Figure~\ref{fig:popA_GG}).
To confirm this, the right panel of Figure~\ref{fig:popA_GG}  
shows that $\rho_{\mA\mA}/\sqrt{S_{\mD}}$  grows linearly\footnote{The 
increase in $\rho_{\mA\mA}/\sqrt{S_{\mD}}$ is no longer linear for large 
values of the Huang-Rhys factors ($S_{\mA}=S_{\mD}\geq1$).} 
with $\Omega_{\mD}$ when $\Omega_{\mA}$ is constant. 
That means, for instance, when $\Omega_{\mA}$ is constant, if $\Omega_{\mD}$ 
increases, then the donor eigenenergies will be higher, since $\mathcal{G}_{\mD}$ is constant, which leads to lower 
donor populations and larger acceptor populations (thermal equilibrium distribution).
Note that $S_{\mD}$ varies inversely with $\Omega_{\mD}$, 
in contrast to the constant Huang-Rhys factors $S_{\mA}=S_{\mD}$ 
analyzed above.
Hence, for constant vibronic coupling strengths 
$\mathcal{G}_{\mA} = \mathcal{G}_{\mD}$,  
$\rho_{\mA\mA}$ becomes larger when 
$\Omega_{\mA} < \Omega_{\mD}$ ($S_{\mA} > S_{\mD}$).

We recently examined the role of the electronic-vibrational (vibronic) 
resonance in the nonequilibrium steady state transport of the PEB dimer in the 
cryptophyte algae PE545 \cite{CCB23}.
We found that an equilibrium steady state reached due to the 
interaction with a thermal bath is not sensitive to vibronic 
resonance.
This is also confirmed for the dimer analyzed here.
When the intramolecular vibrational frequencies at the donor and 
acceptor are in resonance with the excitonic energy splitting 
$\Omega_{\mD},\Omega_{\mA}\approx \Delta E_{e}$  
(i.e., $\Omega_{\mD}=\Omega_{\mA}=142\,\cminv$), the changes 
in the thermal equilibrium acceptor population $\rho_{\mA\mA}$ compared 
to off-resonance frequencies are insignificant ($\lesssim 0.05\%$). 
The analysis and conclusions obtained so far remain valid when 
$\varepsilon_{\mD}<\varepsilon_{\mA}$ with the same 
parameters considered before.

\textbf{\textit{Nonequilibrium steady state configuration}}.\textemdash
The scenario analyzed above, i.e., a thermal equilibrium state reached when  
the system interacts with a vibration bath only, is often examined 
in the context of pulsed laser excitation, in which it is 
assumed, for example, that the coherent light source prepares an initial 
state in the molecular system with the entire population in the donor
or the highest energy exciton state. 
The system then interacts with the vibrational bath and the energy transfer 
dynamics to the acceptor are analyzed \cite{DN&20}.     
However, this differs from natural conditions, in 
which the light-harvesting system is continuously illuminated with incoherent 
natural light \cite{JB91,MV10,BS12,PB12,KYR13,Bru18,PBB17,TK20,CP20,DB21}, 
such as sunlight and additional processes contribute.
Here, we examine the effect of different intramolecular vibrational 
frequencies at the donor and acceptor in the nonequilibrium steady state 
energy transport under the influence of a vibrational bath 
(Eqs.~\ref{eq:HVB}--\ref{eq:SD_DA}), incoherent light excitation, exciton 
recombination, and exciton harvesting at the reaction center. 
We show below that any parameter dependencies observed in the equilibrium 
case for the acceptor population are washed out in the resultant NESS.

The interaction  between the system and the radiation bath 
(incoherent light) is given by the Hamiltonian 
\begin{equation}
\label{eq:Hs-RB}
\hat{H}_{\mathrm{Sys-RB}} = \,
-\hat{\boldsymbol{\mu}}_{\mD}\cdot\hat{\mathbf{E}}(t),
\end{equation}
where $\hat{\boldsymbol{\mu}}_{\mD}$ is the transition dipole operator of 
the donor, and 
$\hat{\mathbf{E}}(t)=\hat{\mathbf{E}}^{(+)}(t)+\hat{\mathbf{E}}^{(-)}(t)$ 
is electric field \cite{MW95}, with 
$\hat{\mathbf{E}}^{(+)}(t)= \mathrm{i}\sum_{\mathbf{k},s}
\sqrt{\frac{\hbar\omega}{2\epsilon_{0}V}}\,\hat{a}_{\mathbf{k},s}\,
\mathbf{e}_{\mathbf{k},s}\mathrm{e}^{-\mathrm{i}\omega t}$
and $\hat{\mathbf{E}}^{(-)}(t)=\left[\hat{\mathbf{E}}^{(+)}(t)\right]^{\dagger}$.
The creation (annihilation) operator for the $\mathbf{k}^{\mathrm{th}}$ 
radiation field mode in the $s^{\mathrm{th}}$ polarization state is denoted 
$\hat{a}_{\mathbf{k},s}^{\dag}$ ($\hat{a}_{\mathbf{k},s}$), and  
$\mathbf{e}_{\mathbf{k},s}$ is the radiation polarization vector.
We assume that the transition electric dipole operator of the acceptor is 
perpendicular to the radiation polarization vector, i.e., only the 
donor chromophore is pumped by incoherent light.
The latter is a typical scenario analyzed in the context of energy transfer
\cite{JB20,JM20a,CCB23}.
The Hamiltonian for the radiation bath reads
\begin{equation}
\label{eq:HRB}
\hat{H}_{\mathrm{RB}}=  
\sum_{\mathbf{k},s}\hbar c \mathrm{k}\,
\hat{a}_{\mathbf{k},s}^{\dag}\hat{a}_{\mathbf{k},s}.
\end{equation}
The radiation bath is described by a super-Ohmic spectral density
with cubic-frequency dependence
\begin{equation}
\label{eq:spectral_density_RB}
J_{\mD}^{\mathrm{\mathrm{RB}}}(\omega)=
\frac{2\hbar\omega^{3}}{3(4\epsilon_{0}\pi^{2}c^{3})}.
\end{equation}
This spectral density generates 
long-lasting coherent dynamics provided by the lack of pure 
dephasing dynamics and by the strong dependence of the decoherence 
rate on the system level spacing \cite{PBB17}.
The temperature assumed for the radiation bath is 
$T^{\mathrm{RB}}=5600\,\mathrm{K}$, and the transition dipole
moment is $\mu_{\mD}=6\,\mathrm{D}$.

We consider the exciton recombination accounting for the nonradiative 
electronic excitation decay to the ground state as a localized process 
on each site and occurring on a nanosecond time scale 
(recombination time $\tau_{\mathrm{rec}}$) 
\cite{MKT14,TB18,JB20,CB20,JM20a}.
The exciton recombination is described by the effective Lindbladian 
\begin{equation}
\label{eq:Lrecloc}
\mathcal{L}_{\mathrm{\mathrm{rec}}}\left[\hat{\rho}\right]=
\tau_{\mathrm{rec}}^{-1}\sum_{i}^{\mD,\mA}
\left(|g_{i}\rangle\langle\varepsilon_{i}|\hat{\rho}|\varepsilon_{i}
\rangle\langle g_{i}|-\frac{1}{2}\left[|\varepsilon_{i}\rangle\langle
\varepsilon_{i}|,\hat{\rho}\right]_{+}\right),
\end{equation}
where $[\hat{\mathcal{O}}_1,\hat{\mathcal{O}}_2]_{+}$ denotes the 
anticommutator  between operators $\hat{\mathcal{O}}_1$ and $\hat{\mathcal{O}}_2$.
The same recombination time for the donor and acceptor is taken as 
$\tau_{\mathrm{rec}} = 1\,\mathrm{ns}$.

The electronic excitation harvesting (trapping) at the reaction 
center occurs on a picosecond time scale 
(trapping time $\tau_{\mathrm{trap}}$). 
Under localized trapping conditions, only the acceptor chromophore is 
coupled to the reaction center, and the trapping process is modeled by the 
Lindbladian \cite{MKT14,TB18,JB20,CB20,JM20a}
\begin{equation}
\label{eq:LRCloc}
\mathcal{L}_{\mathrm{\mathrm{trap}}}^{(\mathrm{loc})}\left[\hat{\rho}\right]=
\tau_{\mathrm{trap}}^{-1}\left(|\mRC\rangle\langle
\varepsilon_{\mA}|\hat{\rho}|\varepsilon_{\mA}\rangle\langle\mRC|
-\frac{1}{2}\left[|\varepsilon_{\mA}\rangle\langle
\varepsilon_{\mA}|,\hat{\rho}\right]_{+}\right).
\end{equation}
For localized trapping conditions and under weak incoherent light excitation 
the NESS quantum yield 
$\eta=\frac{\Gamma_{\mathrm{RC}}}{r_{\mathrm{abs}}}\rho_{AA}^\nequil$ 
is proportional to the acceptor population 
$\rho_{AA}^\nequil$, since the 
ground state population $\rho_{gg}\simeq1$ at all times \cite{TB18,JB20}. 
Here $\Gamma_{\mathrm{RC}} = \tau_{\mathrm{rec}}^{-1}/2$ is the trapping 
rate constant that quantifies the coupling strength to the reaction 
center, and $r_{\mathrm{abs}}$ is the incoherent light absorption rate.

\begin{figure*}[t]
%
\includegraphics[width=0.32\linewidth]{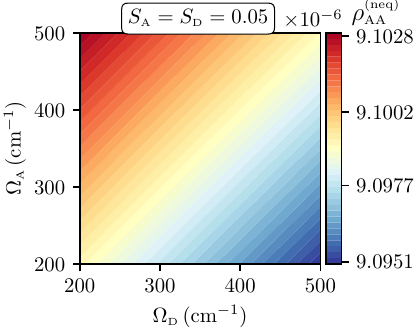} 
\includegraphics[width=0.32\linewidth]{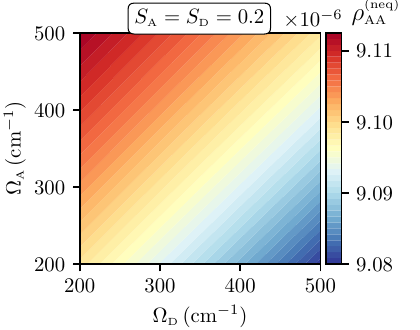} 
\includegraphics[width=0.32\linewidth]{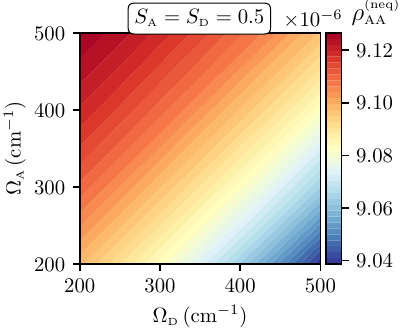} 
\caption{
Nonequilibrium steady state acceptor population $\rho_{\mA\mA}^\nequil$ 
as a function of the intramolecular vibrational frequencies at the donor 
$\Omega_{\mD}$ and the acceptor $\Omega_{\mA}$ considering 
different fixed Huang-Rhys factors $S_{\mA}=S_{\mD}=\{0.05,0.2,0.5\}$ for
localized trapping conditions.
%
The baths parameters are $\lambda=50\,\cminv$, 
$\gamma=200\,\cminv$, $\Gamma=10\,\cminv$, 
$T^{\mathrm{VB}}=300\,\mathrm{K}$, 
$T^{\mathrm{RB}}=5600\,\mathrm{K}$.
Typical values for recombination $\tau_{\mathrm{rec}}=1\,\mathrm{ns}$ and 
trapping $\tau_{\mathrm{trap}}=10\,\mathrm{ps}$ times are assumed.
Note that the color scales  are on the order of $10^{-6}$.
}
\label{fig:popA_NESS_T10}
\end{figure*}

Figure~\ref{fig:popA_NESS_T10} shows the nonequilibrium steady state 
acceptor population $\rho_{\mA\mA}^\nequil$ as a function of the 
intramolecular vibrational donor $\Omega_{\mD}$ and acceptor 
$\Omega_{\mA}$ frequencies for different Huang-Rhys factors
$S_{\mA}=S_{\mD}=\{0.05,0.2,0.5\}$, assuming localized trapping conditions. 
The values reported in color are $10^{-5}$ smaller than those in 
Figure~\ref{fig:popA_SS}, reflecting the weak 
incident incoherent light, exciton harvesting and recombination effects 
\cite{PBB17,CP20,CCB23}. 
We assume a recombination time $\tau_{\mathrm{rec}} = 1\,\mathrm{ns}$ and 
a trapping time $\tau_{\mathrm{trap}} = 10\,\mathrm{ps}$.
The displayed intramolecular vibrational frequencies satisfy 
$\Omega_{\mD},\Omega_{\mA}>\Delta\varepsilon,\Delta E_e,V_{\mD\mA}$.
Figure~\ref{fig:popA_NESS_T10} shows that $\rho_{\mA\mA}^{\nequil}$
increases when the intramolecular vibrational frequency at 
the acceptor is larger than at the donor $\Omega_{\mA}>\Omega_{\mD}$.
Therefore, when $\Omega_{\mA}$ is constant, $\rho_{\mA\mA}^{\nequil}$ 
decreases when $\Omega_{\mD}$ increases.
Figure~\ref{fig:popA_NESS_T10} shows that as $S_{\mA}$ and $S_{\mD}$ increase, 
the maximum and minimum of $\rho_{\mA\mA}^\nequil$ increase and 
decrease, respectively.
The color pattern displayed in Figure~\ref{fig:popA_NESS_T10} is similar to
that of the top panels in Figure~\ref{fig:popA_SS}, for the 
thermal equilibrium state.
However, the changes in $\rho_{\mA\mA}^\nequil$ are seen to be 
insignificant ($<1\%$) for all Huang-Rhys factors considered. 
When the trapping time $\tau_{\mathrm{trap}}$ increases by one order of 
magnitude, e.g., $\tau_{\mathrm{trap}}=100\,\mathrm{ps}$, 
$\rho_{\mA\mA}^\nequil$ also increases by one order of magnitude, and the 
changes the $\rho_{\mA\mA}^\nequil$ are $\sim 3\%$ ($S=0.2$)
(data not shown).
Therefore, increasing the trapping time would enhance the energy transfer 
in the NESS.

In the nonequilibrium steady state, the population flux between the donor and 
acceptor is linked to the imaginary part of the intersite coherence 
\cite{Nit06,RW16,JB20,YC20,CCB23}.
Here, we find that the nonequilibrium steady state intersite coherence 
displays the same trend as for $\rho_{\mA\mA}^{\nequil}$ in 
Figure~\ref{fig:popA_NESS_T10} 
(Figure not shown).
In addition, the case of constant vibronic coupling strengths 
$\mathcal{G}_{\mA}=\mathcal{G}_{\mD}= 100\,\cminv$ was also considered 
with similar results.
The pattern of change in the nonequilibrium steady state acceptor population 
$\rho_{\mA\mA}^\nequil$ is similar to that presented 
for the equilibrium steady state above, 
and the changes $\rho_{\mA\mA}^\nequil$ as those discussed in       
Figure~\ref{fig:popA_NESS_T10}.
Furthermore, even if the reorganization energy increases to 
$\lambda=100\,\cminv$, i.e., twice the value previously considered, and 
still within the relevant range expected under biological conditions, changes 
in the acceptor population are still insignificant (Figure not shown).

\begin{figure*}[t]
\hfill
\includegraphics[width=0.32\linewidth]{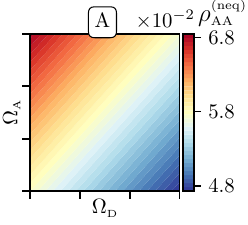}
\hspace{-2mm}
\includegraphics[width=0.32\linewidth]{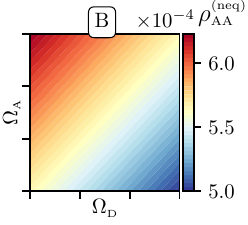}
\hspace{-2mm}
\includegraphics[width=0.32\linewidth]{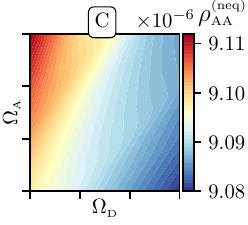}
\hspace{-1mm}
\includegraphics[scale=0.9]{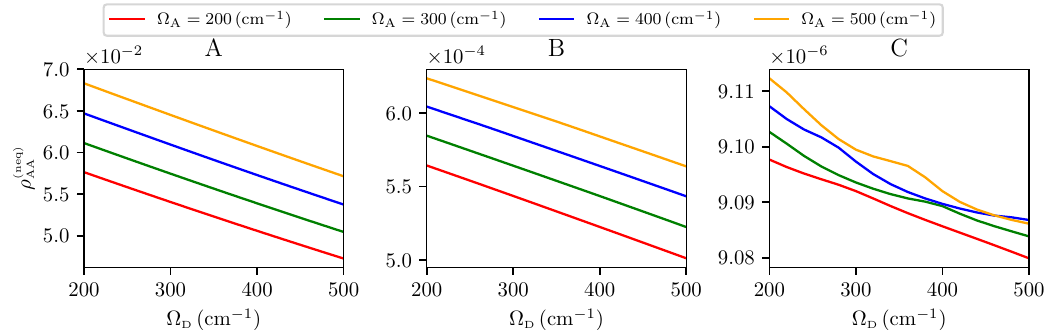}
\hfill
\caption{
Nonequilibrium steady state acceptor population $\rho_{\mA\mA}^\nequil$ as a 
function of the intramolecular vibrational frequency at the donor 
$\Omega_{\mD}$ and acceptor frequencies $\Omega_{\mA}$.
We consider different scenarios for the NESS, radiation and vibrational baths 
only (panel $\mathrm{A}$), radiation bath and exciton recombination only 
(panel $\mathrm{B}$), radiation bath, and exciton recombination and localized 
exciton trapping (panel $\mathrm{C}$).
We assume the Huang-Rhys factors $S=S_{\mA}=S_{\mD}=0.2$.
The vibrational bath parameters are $\lambda=50\,\cminv$, 
$\gamma=200\,\cminv$, $\Gamma=10\,\cminv$ , and 
$T^{\mathrm{VB}}=300\,\mathrm{K}$.
Typical values for recombination $\tau_{\mathrm{rec}}=1\,\mathrm{ns}$ and 
trapping $\tau_{\mathrm{trap}}=10\,\mathrm{ps}$ times are assumed.
Note the differences in the vertical scale.
}
\label{fig:popA_NESS_nonunitary_effects}
\end{figure*}

%
Even though the natural scenario approximates the picture discussed above, 
where the exciton population is collected at the reaction center and 
excitons have a finite lifetime (exciton recombination), we examine  
individual non-unitary contributions to evaluate the effect of the 
donor-acceptor intramolecular vibrational frequency difference in the NESS 
acceptor population.
Figure~\ref{fig:popA_NESS_nonunitary_effects} depicts the acceptor population 
$\rho_{\mA\mA}^\nequil$  as a function of $\Omega_{\mD}$ and $\Omega_{\mA}$ 
for a fixed Huang-Rhys factor $S_{\mA}=S_{\mD}=0.2$, considering that the 
system only interacts with: the incoherent 
light and the vibrational bath (panel $\mathrm{A}$), the incoherent light 
and also undergoes exciton recombination (panel $\mathrm{B}$), and the 
incoherent light and also undergoes exciton recombination and localized trapping 
(panel $\mathrm{C}$).
In all the cases, $\rho_{\mA\mA}^\nequil$ is larger when 
$\Omega_{\mA}>\Omega_{\mD}$ and decreases as $\Omega_{\mD}$ increases.
Figure~\ref{fig:popA_NESS_nonunitary_effects} shows that $\rho_{\mA\mA}^\nequil$ 
decreases linearly with $\Omega_{\mD}$ for panels $\mathrm{A}$ and $\mathrm{B}$ 
(no exciton harvesting), but when the effect of exciton trapping is included, 
the decrease is no longer linear.
Therefore, the nonlinear variation of $\rho_{\mA\mA}^\nequil$ with 
$\Omega_{\mD}$ is caused by exciton trapping at the reaction center, 
and increases when the trapping time $\tau_{\mathrm{trap}}$ decreases. 
In addition, as discussed above for the equilibrium case, the nonlinearity 
also increases when the Huang-Rhys factors increase. 
It is important to note that when the system interacts with the incoherent 
light and the vibrational bath only, the NESS reached allows for significant 
changes up to $40\%$ in $\rho_{\mA\mA}^\nequil$ (panel $\mathrm{A}$). 
Under this scenario the interaction with the vibrational bath is stronger than 
the interaction with incoherent light.
Therefore, it is the exciton harvesting at the reaction center that washes 
out the variations in the acceptor population related to the donor-acceptor 
intramolecular vibrational frequency difference displayed in the thermal
equilibrium case.

In conclusion, the HEOM method was used to analyze the effect of different 
intramolecular vibrational frequencies on energy transfer in a prototype 
photosynthetic dimer system, under both equilibrium (ESS) and nonequilibrium 
steady state (NESS) conditions. 
When an equilibrium thermal state is reached by the interaction of the 
system only with a vibrational bath, the results indicate 
that, for constant Huang-Rhys factors, the thermal equilibrium acceptor 
population (quantum yield)  decreases with an increasing vibrational donor 
frequency for a constant vibrational acceptor frequency. 
That is, the acceptor population increases when the vibrational 
frequency of the acceptor is larger than in the donor because the vibronic 
coupling in the acceptor is larger than in the donor.
Conversely, for constant vibronic coupling strengths, the acceptor population 
increases proportionally to the square root of the donor 
Huang-Rhys factor times the vibrational donor frequency
for a constant vibrational acceptor frequency.
In this case, the quantum yield increases when the vibrational frequency 
of the donor is larger than in the acceptor because the Huang-Rhys 
factor in the acceptor is larger than in the donor.

In the NESS, and considering constant Huang-Rhys factors, the variations in 
the acceptor population for localized trapping conditions are much smaller 
than those of the thermal equilibrium quantum yield.
Therefore, under natural biological conditions of incoherent light excitation, 
the NESS reached does not allow for a significant enhancement in the quantum 
yield due to the intramolecular vibrational frequency difference.
Technologically, however, one can increase the quantum yield somewhat by 
increasing the trapping time. 
For example, increasing the trapping time by one order of magnitude increases
the acceptor population by  one order of magnitude, with a higher variation 
with the donor-acceptor intramolecular vibrational frequency difference.
Therefore, increasing the trapping time is a possible mechanism to enhance 
the energy transfer in the NESS.  
In the future, it would be interesting to examine more realistic exciton 
trapping scenarios beyond the effective Lindblad methodology considered in this 
Letter.

\begin{acknowledgement}
This work was supported by the U.S. Air Force Oﬃce of Scientific  
Research (AFOSR) under Contract No. FA9550-20-1-0354.
\end{acknowledgement}

\bibliography{refs}

\end{document}